# The mechanism of Supernova Ia explosion in elliptical galaxies

V M Lipunov[1], I E Panchenko[1] and M V Pruzhinskaya[1]

[1]*Sternberg Astronomical Institute, Moscow State University, 119992, Moscow, Russia;*

E-mail: lipunov@sai.msu.ru

**Abstract.** Recent observational data on the type Ia supernova rates are in excellent agreement with the earlier results of the population synthesis of binary stars and confirm that the overwhelming majority of type Ia supernovas (~99%) in elliptical galaxies form via mergers of binary white dwarfs with a total mass exceeding the Chandrasekhar limit.

## 1. Introduction

The interest in type Ia supernovas (used as standard candles for cosmology), which led the researchers to suspect the presence of dark energy in the Universe (Riess et al. 1998; Perlmutter et al. 1999), triggered mass discovery of supernovas, resulting in an almost 50-fold increase of the number of these stars studied in the last decade! The mass discovery of supernovas in recent years allowed the researchers to observe for the first time the dramatic evolution of the supernova rate in elliptical galaxies, which was predicted more than 10 years ago via population synthesis (Jorgensen et al. 1997).

Type Ia supernovas are now generally believed to be products of nuclear explosions of white dwarfs that have reached the Chandrasekhar limit (for a review see Livio et al. 2000).

No appreciable star formation goes on in elliptical galaxies. Only low-mass stars remain in these systems after the first billion years of their evolution. The evolution of all massive stars (with $M > 8\text{-}10\ M_{\odot}$) ends completely with the formation of neutron stars and black holes. Low-mass stars by themselves cannot produce supernova explosions, because their evolution ends with a soft formation of white dwarfs with masses below the stability limit (the Chandrasekhar limit). However, a delayed (by several billion years) accumulation of the Chandrasekhar mass may occur in binary systems – as a result of either the accretion of matter from a companion (the so-called SD-mechanism (Whelan & Iben 1973)), or merger (the DD-mechanism (Iben & Tutukov 1984; Webbink 1984)).

## 2. Scenario Machine Prediction and recent observations

The very first evolutionary computations of such processes in elliptical galaxies (population synthesis) performed using a special computer code – the Scenario Machine (Lipunov et al. 1996; Lipunov et al. 2009) – showed (Jorgensen et al. 1997) that the mechanism of white-dwarf merger outperforms accretion by two orders of magnitude already one billion years after the formation of the elliptical galaxy (see Figure 1).

# The mechanism of Supernova Ia explosion in elliptical galaxies

The study of supernovas in recent years allowed the evolution of the supernova rates in elliptical galaxies to be observed for the first time (Totani et al. 2008) (Figure 1). These results were obtained by analyzing the observations of candidate type Ia supernovas based on Subaru/XMM-Newton Deep Survey (SXDS) data. The ages of elliptical galaxies were determined from nine-band photometry spanning from optical to mid-infrared wavelengths. The observed decrease of the SN Ia rate was found to be described by the $f \sim t^{-\alpha}$ law, where $\alpha \approx 1$.

Supernova observers measure the supernova rates per unit *K*-band absolute magnitude $10^{10} L_{K,0,\odot}$ and we therefore converted our old data into the new plot assuming that $M_{*}/L_{K,0}$=1.8[$M_{\odot}/L_{K,\odot}$] in accordance with modern data to find the results to be in excellent agreement (Figure 1) with the supernova rates predicted by the theory of the evolution of binary stars (Jorgensen et al. 1997) .

It goes without saying that population synthesis is a complex numerical process, which incorporates our knowledge and hypotheses about the evolution of binary stars, as well as the observed properties of binary stars (the initial mass function and the initial distribution of separations). However, we try to show that the results obtained 13 years ago, like those of more recent computations (Fedorova et al. 2004; Förster et al. 2006; Totani et al. 2008; Wang et al. 2010; Yungelson et al. 1996; Yungelson 2005), are extremely weakly sensitive to the «dark areas» of the evolution of binary stars.

Here the form of the initial distribution of separations of binary systems plays the crucial part (a).

It was shown (Popova et al 1982; Abt 1983) that the observed distribution of separations of binary stars in our Galaxy at the beginning of their main-sequence evolution can be described by the following law:

$$dN \sim \varphi(a)da \sim \frac{1}{a}da \qquad (1)$$

This distribution still remains a theoretical puzzle, which can be popularly formulated as follows – our Galaxy contains approximately equal numbers of wide and close binaries (i.e., equal logarithmic intervals – e.g., the decades – contain equal numbers of stars). We can assume, to a first approximation (as it is commonly done in population synthesis), that other galaxies must have had the same initial distribution of binary stars. There are no particular reasons to believe that binaries in other galaxies should form in a different way. After its formation a binary star undergoes a long and varied evolution accompanied by the change of the component separation. In low-mass binaries the most important and least understood evolutionary factor remains the so-called common-envelope phase, where one of the components is inside its companion star swollen to the red—giant state (stellar cannibalism). During the common-envelope stage the components approach each other catastrophically. However, because of the power-law form of distribution (1), a proportional approach by a certain factor that is almost independent of the component separation has no appreciable effect on the distribution function. Hence it would appear logical to suggest that at the time of the formation of binary white dwarfs the distribution function of their separations can still be described by the flat law (1).

The subsequent evolution of each binary consists in the two white dwarfs slowly approaching each other because of the emission of gravitational waves. This approach can be described by the following Einstein's formula (Landau & Lifshitz 1975):

$$\frac{da}{dt} \sim a^{-3}(M_1 + M_2)M_1 M_2 \qquad (2)$$

where $M_1$ and $M_2$ are the component masses, which we assume to be equal. The time scale of the evolution of a star until it becomes a white dwarf is determined mostly by the hydrogen burning time $T \sim 10^{10}(M/M_{\odot})^{-2}$ *yrs* and hence the merger time scale should be determined by some power law.

We now demonstrate that the change of the supernova rate due to DD mechanism is independent of the component approach law!

Indeed, the white-dwarf merger rate is proportional to the number of systems with the given separation and the rate of the decrease of the component separation:

$$\frac{da}{dt} \sim a^n \rightarrow t \sim a^{-n+1} \tag{3}$$

Let $\varphi \sim a^{-\beta}$

The Merging Rate is the

$$Rate \sim \varphi \frac{da}{dt} \sim a^{-\beta} a^n \sim t^{-(\frac{n-\beta}{n-1})} \tag{4}$$

It is remarkable that if $\beta = 1$ (see equation (1)) the result does not depend on a particular merger mechanism (cf. Totani et al. 2008) – it is only important that the merger time scale should be determined by the initial separation. This is especially important, because the formation of white dwarfs is not an instantaneous process. To a first approximation, the lifetime of a star of mass $M$ is proportional to $M^{-2}$ and hence new binary white dwarfs of increasingly smaller masses should form in an elliptical galaxy after several billion years. However, their semimajor axes will always be distributed in accordance with law (1) and law (2) would remain unchanged.

Thus the observed variation of the supernova rate in elliptical galaxies confirms not only the model of white-dwarf mergers as the main mechanism of type Ia supernovae explosions. However at early time ($\sim 10^8$ yrs) accretion mechanism is the main (Yungelson 2010) and we predict a high soft X-ray luminosity from galaxies with this age.

## 3. Summary and discussions

We finally point out a recent paper by Gilfanov and Bogdan (2010), who, based on completely different considerations, conclude that DD-process should be the dominating mechanism of type Ia supernova formation in elliptical galaxies. The above authors proceed from a simple logic that for the SD mechanism to operate, constant accretion is needed, which must be accompanied by soft x-ray flux, which, in turn, are found to be are smaller than predicted for the mechanism considered. However, first, the absence of something cannot be taken for positive evidence (as Ya. B. Zel'dovich used to say, the fact that no electric wires have been found in excavation sites in Rome does not prove that ancient Romans had radio), and, second, it is not the accretion rate, but rather the integrated mass (over time) that is really important for the SD mechanism. Hence the x-ray luminosity should not be directly related to the supernova rate. In elliptical galaxies white dwarfs may grow their mass in systems of three types: (1) if a red- or yellow-dwarf secondary fills its Roche lobe (astronomers observe such systems as cataclysmic variables) or (2) if a helium secondary fills its Roche lobe (3) in wide systems where the white-dwarf's companion is a red giant (symbiotic stars). In the former case theoretical computations show that the white dwarf may grow to Chandrasekhar mass only if the accretion rate exceeds $10^{-7}$ solar masses/years, otherwise the hydrogen-and-helium mix will not burn up completely during the accretion, but will rather be ejected during recurrent nova outbursts. The above authors use this latter circumstance to predict the x-ray luminosities of elliptical galaxies. However, no one knows the fraction of accumulated mass that after the nova explosion remains on the white dwarf surface and pushes it toward the Chandrasekhar limit. For example, a 5-10% or higher fraction is sufficient for heavy dwarfs to become unstable. In this case the x-ray luminosity must be very low. In the case of symbiotic binaries the soft x-ray radiation must be absorbed entirely in the outflowing stellar wind of the red giant. Thus a discussion of the role of absorption in the envelopes of symbiotic stars and helium stars, x-ray spectra, and stationarity of burning in the accretion flow shows how long is the way that theory has to go before the lack of x-ray radiation becomes a compelling argument.

**Acknowledgements** The authors are grateful to Alexander Tutukov and Evgeny Gorbovskoy for helpful discussions.

This work was supported by the Ministry of Science of the Russian Federation (state contract N02.740.11.0249).

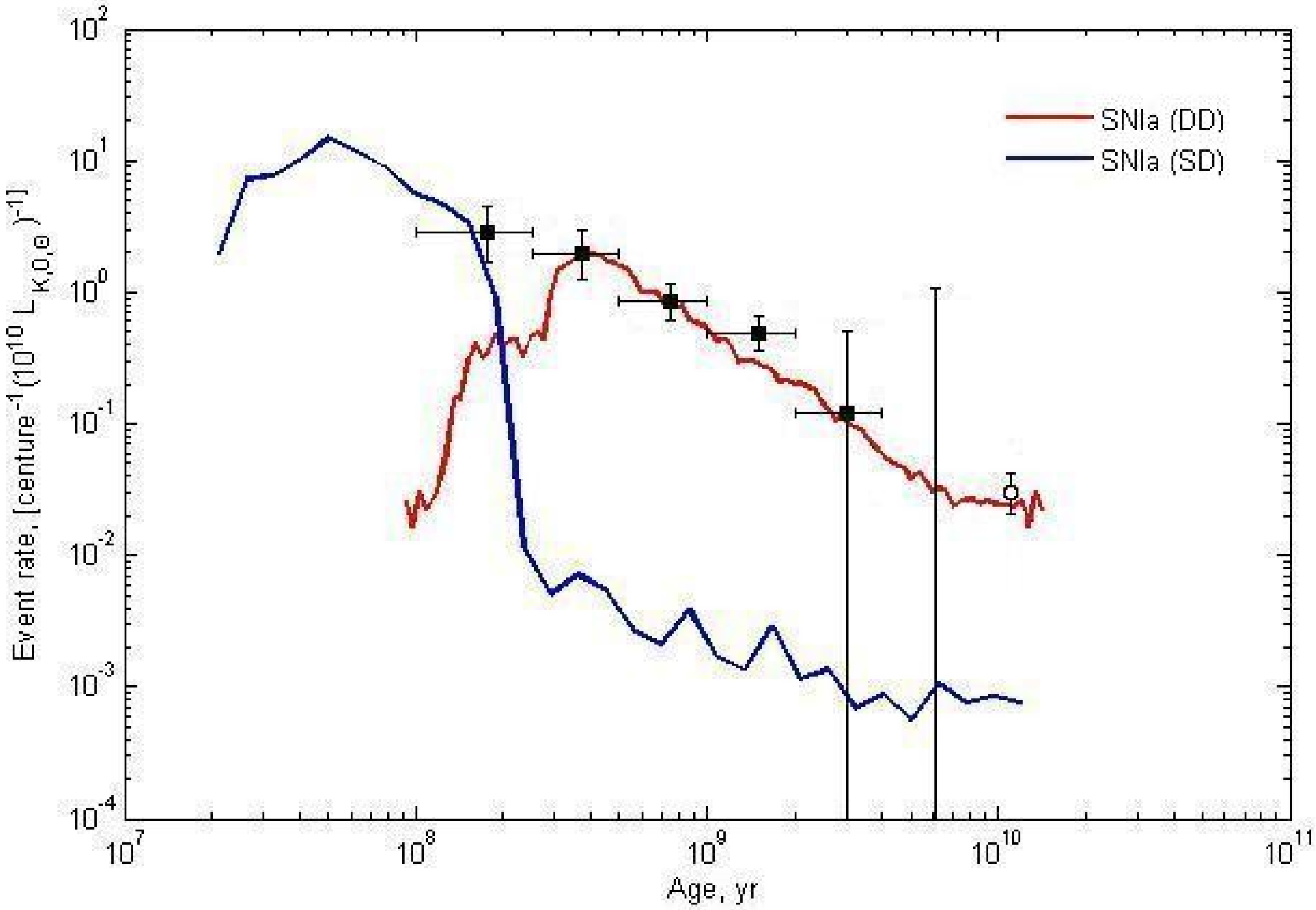

**Figure 1.** The SN Ia rate per century for a single starburst population whose total $K$-band luminosity is $10^{10}L_{K,\odot}$ at the age of 11 Gyr [centure$^{-1}$($10^{10}$ $L_{K,0,\odot}$)$^{-1}$] (predicted by Jorgensen et al. 1997). The filled squares are the observational points[9]. The open circle is observational SN Ia rate in elliptical galaxies in the local Universe (Mannuci et al. 2006).

**References**

Abt, H. A. 1983, ARA&A, **21**, 343

Fedorova, A.V., et al. 2004, Astronomy Letters, **30**, 73

Förster, F., Wolf, C., Podsiadlowski, P. & Han, Z. 2006, MNRAS, **368**, 1893

Gilfanov, M., & Borgan, A. 2010, Nature, **463**, 924

Iben, I. Jr., & Tutukov, A.V. 1984, ApJS, **54**, 335

Jorgensen, H. E., Lipunov, V. M., Panchenko, I. E., Postnov, K. A., & Prokhorov, M. E. 1997, ApJ, **486**, 110

Landau, L. D., & Lifshitz, E. M. 1975, The Classical Theory of Fields (4th ed.; Elsevier, Oxford)

Lipunov, V. M., Postnov, K. A., & Prokhorov, M. E. 1996, Astrophysics and Space Physics Reviews, Vol. 17, ed. R.A. Sunyaev (Harwood Acad. Publ.)

Lipunov, V. M., Postnov, K. A., Prokhorov, M. E., & Bogomazov, A. I. 2009, ARep, **53**, 915

Livio, M. 2000, ed. J.C. Niemeyer, & J.W. Truran (Cambridge, Cambridge University Press), **33**

Mannucci, F., et al. 2005, A&A, **433**, 807

Perlmutter, S., et al. 1999, ApJ, **517**, 565

Popova, E.I., Tutukov, A.V., & Yungelson, L.R. 1982, Ap&SS, **88**, 55

Riess, A.G., et al. 1988, ApJ, **116**, 1009

Totani, T., Morokuma, T., Oda, T., et al. 2008, PASJ, **60**, 1327

Wang, B., Liu, Z., Han, Y., Lei, Z., Luo, Y. & Han, Z. 2010, ScChG (Sci. China Ser. G), 53, 586

Webbink, R. 1984, ApJ, **277**, 355

Whelan, J., & Iben, I.Jr. 1973, ApJ, **186**, 1007

Yungelson, L.R., et al. 1996, ApJ, **466**, 890

Yungelson, L.R. 2005, ASSL, **332**, 163

Yungelson, L.R. 2010, Astronomy Letters, **36**, 780